\documentclass[11pt,letterpaper]{article}

\usepackage[utf8]{inputenc}
\usepackage[T1]{fontenc}
\usepackage{amsmath, amssymb, amsfonts}
\usepackage{graphicx}
\usepackage{booktabs}
\usepackage{float}
\usepackage{authblk}
\usepackage{setspace}
\usepackage{geometry}
\usepackage{caption}
\usepackage{subcaption}
\usepackage[hidelinks]{hyperref}
\usepackage[numbers,sort&compress]{natbib}
\usepackage[hang,flushmargin]{footmisc}
\usepackage{fancyhdr}
\title{\vspace{200pt} \textbf{Deep Learning Framework for RNA Inverse Folding with Geometric Structure Potentials}}
\author[a]{\vspace{30pt} \Large Annabelle Yao}
\date{} 

\begin{document}

\maketitle
\thispagestyle{empty} 

\newpage
\thispagestyle{empty} 
\begin{abstract}
\noindent RNA’s diverse biological functions stem from its structural versatility, yet accurately predicting and designing RNA sequences given a 3D conformation (inverse folding) remains a challenge. Here, I introduce a deep learning framework that integrates Geometric Vector Perceptron (GVP) layers with a Transformer architecture to enable end-to-end RNA design. I construct a dataset consisting of experimentally solved RNA 3D structures, filtered and deduplicated from the BGSU RNA list, and evaluate the performance using both sequence recovery rate and TM-score to assess sequence and structural fidelity, respectively. On standard benchmarks and RNA-Puzzles, my model achieves state-of-the-art performance, with recovery and TM-scores of 0.481 and 0.332, surpassing existing methods across diverse RNA families and length scales. Masked family-level validation using Rfam annotations confirms strong generalization beyond seen families. Furthermore, inverse-folded sequences, when refolded using AlphaFold3, closely resemble native structures, highlighting the critical role of geometric features captured by GVP layers in enhancing Transformer-based RNA design.\\
\end{abstract}

\newpage

\maketitle

\newpage
\tableofcontents\thispagestyle{empty} 
\newpage
\setcounter{page}{1}
\section{Introduction}

RNA is a highly versatile biomolecule that plays essential roles in a broad spectrum of biological processes. Beyond serving as a messenger between DNA and proteins, RNA is central to gene regulation, catalysis, and cellular signaling. Its primary structure comprises four nucleotides—adenine (A), cytosine (C), guanine (G), and uracil (U)—and RNA molecules are generally classified into two major types: coding RNAs (mRNAs), which direct protein synthesis, and non-coding RNAs (ncRNAs), which function primarily through their intricate three-dimensional (3D) structures. The folding of RNA into these structures is largely driven by base-pairing interactions at the secondary structure level. This tight coupling between structure and function underscores the importance of accurately characterizing RNA 3D conformations, which is key to elucidating the molecular basis of many diseases and fundamental cellular processes.
 
Although understanding RNA structure is of profound biological significance, the conformational flexibility of RNA molecules poses a major challenge for experimental determination of their three-dimensional (3D) structures. As of December 2023, RNA-only structures constitute less than 1.0\% of the approximately 214,000 entries in the Protein Data Bank (PDB), while RNA-containing complexes account for just 2.1\%. This scarcity reflects the intrinsic difficulty of resolving RNA structures, which often adopt a diverse ensemble of conformations due to their dynamic nature. Despite advances in traditional experimental techniques, including X-ray crystallography, nuclear magnetic resonance (NMR) spectroscopy, and cryogenic electron microscopy (cryo-EM), these methods remain limited by low throughput, labor-intensive sample preparation, and specialized equipment requirements. Consequently, the large-scale elucidation of RNA structures remains impractical through experimental means alone. To address these limitations, computational approaches have emerged as essential tools for RNA structure prediction, offering scalable and efficient alternatives to experimental methods. The recent rise of artificial intelligence (AI) has further revolutionized the field, enabling significant gains in prediction accuracy and structural insight. In particular, the 2024 Nobel Prize in Chemistry awarded to the developers of AlphaFold \cite{dauparas2022robust} exemplifies the transformative power of AI in structural biology. Simultaneously, we have seen breakthroughs in RNA structure prediction, with deep learning-based methods such as Rhofold+ \cite{shen2024accurate}, trRosettaRNA \cite{wang2023trrosettarna}, and others demonstrating remarkable improvements over traditional approaches like FARFAR2 \cite{watkins2020farfar2} and SimRNA \cite{boniecki2016simrna}. 

Another crucial aspect of RNA study lies in inverse folding, which has gained increasing significance with advances in RNA structure prediction. While elucidating the relationship between RNA structure and function remains fundamental to understanding its biological roles, inverse folding takes this one step further by seeking RNA sequences that adopt a predefined 3D conformation. This capability is essential for the rational design of synthetic RNA molecules with desired structural and functional properties. Applications span a wide range of fields, including RNA-based therapeutics, gene regulation, biosensing, and synthetic biology. Moreover, accurate inverse folding not only facilitates the engineering of functional RNA elements but also enables the discovery of novel structural motifs involved in critical cellular processes. By bridging the gap between structure prediction and functional design, inverse folding empowers researchers to manipulate RNA with precision, paving the way for breakthroughs in biotechnology, medicine, and the development of RNA-based nanodevices. 

Despite the growing potential of RNA inverse folding, designing sequences that reliably adopt a target 3D structure remains a formidable challenge. This difficulty stems from the enormous combinatorial space of nucleotide sequences and the complex network of physical and chemical interactions that govern RNA folding. Traditionally, researchers have addressed this problem through methods like RNA Origami, which involves engineering nanostructures that self-assemble co-transcriptionally with the aid of software such as NUPACK \cite{zadeh2011nupack} and ViennaRNA \cite{lorenz2011viennarna}. These methods have proven effective for in vitro applications and have supported the development of increasingly complex RNA architectures. However, they are often limited by their reliance on secondary structure approximations and may not generalize well to the full three-dimensional complexity required for functional RNA design in more diverse biological contexts. 

The potential of RNA inverse folding extends far beyond the scope of traditional computational strategies. Recent advances in deep learning have introduced powerful new approaches for navigating the immense complexity of RNA folding. While protein inverse folding has seen rapid progress through groundbreaking models such as AlphaFold and ProteinMPNN, RNA presents unique challenges, including non-canonical base-pairing, intricate tertiary interactions, and highly diverse structural motifs. These complexities have limited the widespread adoption of deep learning for RNA inverse folding. As a result, fewer studies have applied deep learning in this domain, with notable exceptions including Monte Carlo tree search–based frameworks \cite{yang2017rna} and the generative diffusion model RiboDiffusion \cite{huang2024ribodiffusion}. In addition, successful applications of machine learning, such as transformers in computer vision \cite{yao2024enhancing}, have shown that deep learning models work best when sharing features with the domain's problem structure. Building upon these insights, this work proposes a deep learning approach that integrates geometric considerations directly into the model architecture by leveraging a Geometric Vector Perceptron (GVP) with a Transformer to address a new significant problem within the RNA field. \cite{jing2020learning}. My approach achieves substantially improved performance on both standard benchmarks and RNA-Puzzles, outperforming existing methods in terms of sequence recovery and structural fidelity, and highlighting its effectiveness in accurate RNA sequence design and 3D structure prediction.

\section{Methods}

\subsection{Data Preprocessing} 
The dataset used for this study and evaluation was obtained from the BGSU RNA list version 3.321, which contains a collection of experimentally solved RNA 3D structures. The initial dataset included over 6300 RNA 3D structures, downloaded from the RCSB Protein Data Bank (PDB). To preprocess the dataset, the RNA sequences were first filtered to retain only those with lengths between 20 and 512 nucleotides. Following this, single-chain RNA sequences were extracted from the resulting filtered 3D structures. Redundancy was then removed using the CD-Hit tool with an 80\% sequence similarity cutoff, resulting in approximately 2500 unique single-chain RNA 3D structures. For each nucleotide in these sequences, key atoms (P', C5', O5', C4', O4', C3', O3', C2', O2', C1', N1, C2, O2, N3, C4, N4, C5, C6, OP1, and OP2), which are essential for the model, were extracted. Any missing atoms were filled in using Arena \cite{perry2023arena}. The sequence of these single chains formed the desired output, which was used for both structural and sequence comparisons. This pre-processing pipeline ultimately produced a dataset of around 2500 RNA single-chain 3D structures, which was utilized for evaluating the recovery rate and structural prediction. A diagram of my data pre-processing method is shown below in Figure~\ref{fig:Datapreprocessing}.

\begin{figure}[!ht]

     \makebox[\textwidth]{%
        \includegraphics[width=0.95\textwidth]{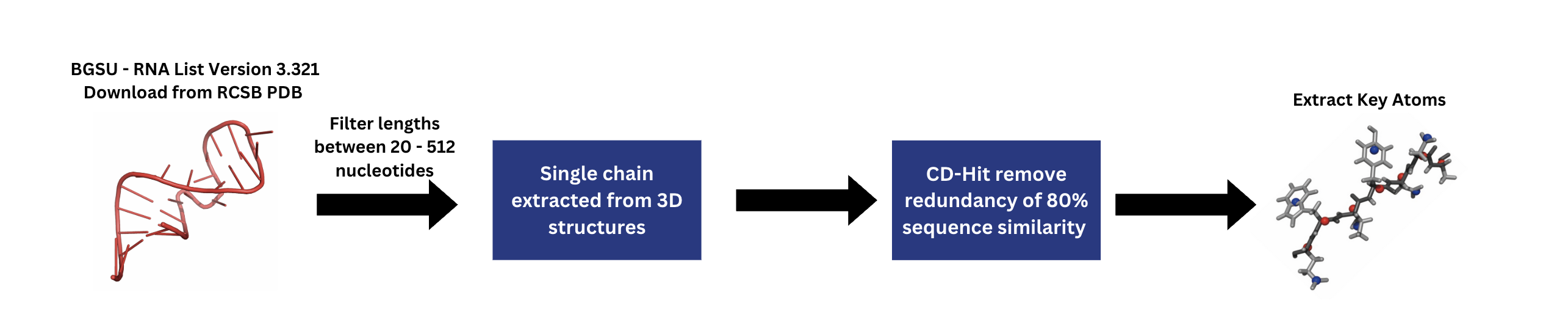}
    }
    \caption{The overall method used to preprocess my data. (Graph created by the student researcher using Canva and the structures downloaded from RCSB PDB)}
    \label{fig:Datapreprocessing}
\end{figure}

\begin{figure}[!ht]

     \makebox[\textwidth]{%
        \includegraphics[width=1.1\textwidth]{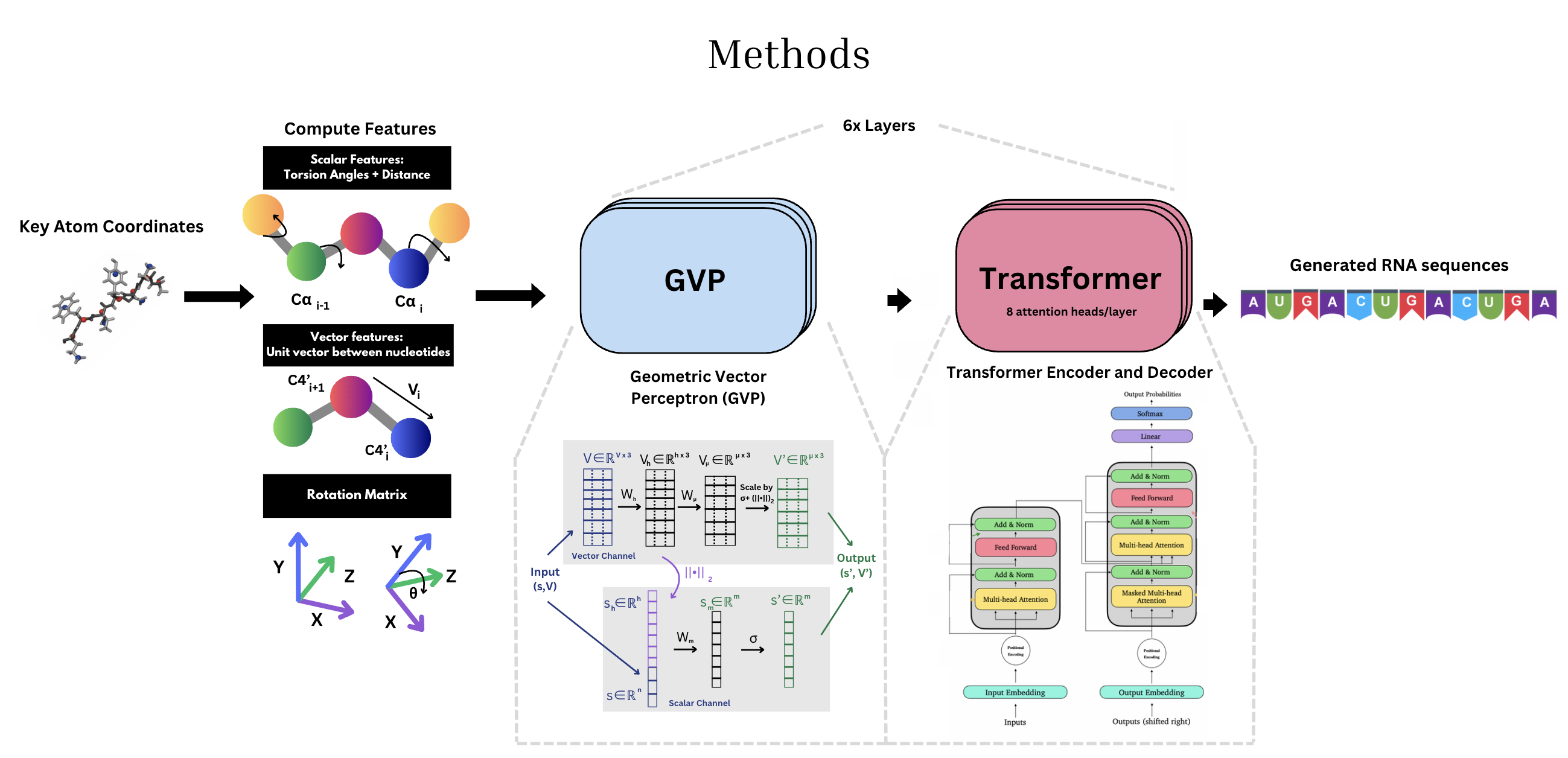}
    }
    \caption{The workflow of my framework. First, key atom coordinates are extracted from the PDB file. Then, the scalar and vector features of the key atoms are computed along with the rotation matrix to ensure rotational invariance. These features will be used to make a graph where the nodes are each key atom and the weights are the features. The graph is then passed through the GVP and the Transformer to generate an RNA sequence. (Graph created by the student researcher using Canva and the structures downloaded from RCSB PDB)}
    \label{fig:OverallMethods}
\end{figure}

Given an RNA tertiary structure, I extracted its RNA sequence along with the coordinates of the key atoms for each nucleotide. The key atoms include: P', C5', O5', C4', O4', C3', O3', C2', O2', C1', N1, C2, O2, N3, C4, N4, C5, C6, OP1, and OP2. Based on this information, I frame my RNA inverse folding problem: Given the coordinates of these key atoms, the goal is to generate RNA sequences that fold into the specified structure. To solve this problem, my aim was to construct a deep learning model capable of predicting RNA sequences that correspond to a given structural configuration. I used a Geometric Vector Perceptron (GVP) and a transformer to generate RNA sequences. The general method is shown in Figure~\ref{fig:OverallMethods}.

\subsection{Structural Features} After extracting the coordinates of the key atoms from the PDB file, I calculate the torsion angles and the vectors between the coordinates. Torsion angles are the angles between two planes formed by specific key atoms and describe the flexibility of the backbone, playing a crucial role in defining an RNA's 3D structure. RNA has six backbone torsion angles per nucleotide, defined around the covalent bonds, the glycosidic bond, and the sugar pucker. These angles are measured around the bonds connecting different backbone atoms as below. Each angle is calculated by finding the difference between four points p0 - p1 - p2 - p3. Then, the difference between p2 and p1, the unit vector b1 along the axis of rotation, is normalized to ensure that it has a magnitude of 1. I calculate the orthogonal vectors by removing the component along b1 using the dot product. These vectors are then used to calculate the torsion angles with the equations below, with (i-1) and (i+1) indicating atoms from adjacent nucleotides in the RNA chain.

\begin{itemize}
    \item $\alpha$ (alpha): O3' (i-1) $\to$ P $\to$ O5' $\to$ C5'
    \item $\beta$ (beta): P $\to$ O5' $\to$ C5' $\to$ C4'
    \item $\gamma$ (gamma): O5' $\to$ C5' $\to$ C4' $\to$ C3'
    \item $\delta$ (delta): C5' $\to$ C4' $\to$ C3' $\to$ O3'
    \item $\epsilon$ (epsilon): C4' $\to$ C3' $\to$ O3' $\to$ P (i+1)
    \item $\zeta$ (zeta): C3' $\to$ O3' $\to$ P (i+1) $\to$ O5' (i+1)
\end{itemize}

These angles serve as the scalar features of each nucleotide. The vector values for each nucleotide are then calculated by finding the displacement vector, a 3D spatial vector, between the C4' atoms of nucleotide i and nucleotide i+1. The scalar and vector features are stored in matrices  \( s\) and  \( V \) where \begin{align*}
    s &\in \mathbb{R}^{N \times d_s} \\
    V &\in \mathbb{R}^{N \times d_v \times 3}
\end{align*}
where \( N \) is the number of nucleotides, \( d_s \) is the number of scalar features, and \( d_v \) is the number of vector features.

Moreover, to ensure that the model is learning rotation-invariant features, I extract 3 atoms per nucleotide, specifically N1, C4', and C1', to construct a local coordinate frame. I use these atoms to compute a rotation matrix $\mathbb{R}$ for each nucleotide, ensuring that all vector features are expressed in a consistent local frame. First, I compute the Euclidean norm that I call L2 along a given dimension. The three orthogonal base vectors are calculated through these equations:
\begin{align}
    \mathbf{e}_1 &= \text{normalize}(\mathbf{v}_1, \text{dim}=-1) \quad \text{(unit vector along C $\to$ CA)} \\
    \mathbf{e}_2 &= \text{normalize}(\mathbf{u}_2, \text{dim}=-1) \quad \text{(adjusted vector orthogonalized against $\mathbf{e}_1$)} \\
    \mathbf{e}_3 &= \mathbf{e}_1 \times \mathbf{e}_2 \quad \text{(cross product ensuring a right-handed coordinate system)}
\end{align}
These vectors form the rotation matrix:
\begin{equation}
    R = \begin{bmatrix} \mathbf{e}_1 & \mathbf{e}_2 & \mathbf{e}_3 \end{bmatrix}
\end{equation}
which defines the local coordinate frame.
After I generated these structural features, I applied an Encoder-Decoder-based network to process my input.

\subsection{Feature Encoding}
I use a model architecture combining Geometric Vector Perception (GVP) and a Transformer encoder framework to encode RNA structural information. The RNA structure is represented as a graph, where each nucleotide is treated as a node and the node feature is calculated using torsion angles and rotation matrices. The edges between nodes are determined by the spatial proximity of nucleotides, calculated through pairwise distances. To construct this graph, I calculate a distance matrix based on the coordinates of the C4' atoms of each nucleotide, which serves as a reference for the relative positioning of nucleotides. From this matrix, I compute the Euclidean distances between all pairs of nucleotides, storing these distances in another matrix.

Using the calculated distance matrix, I identify the five nearest neighbors for each nucleotide and establish edges between the corresponding nodes. This process ensures that only relevant, short-range connections are included in the graph, representing the structural relationships between adjacent nucleotides. The RNA molecule is now fully encoded as a graph, with nodes (nucleotides), edges (connections between neighboring nucleotides), and node features (scalar and vector features calculated from the structural data).

The constructed graph is then fed into a Geometric Vector Perception (GVP) layer. GVP is a powerful module designed to learn vector- and scalar-valued functions over geometric vectors and scalars. Given a tuple $(\mathbf{s}, \mathbf{V})$ of scalar features $\mathbf{s} \in \mathbb{R}^n$ and vector features $\mathbf{V} \in \mathbb{R}^{\nu \times 3}$, the GVP module computes new features $(s', \mathbf{V'}) \in \mathbb{R}^{m} \times \mathbb{R}^{\mu \times 3}$. The overall diagram of a GVP is shown below in Figure~\ref{fig:GVP-GCN}.

\begin{figure}[!ht]

     \makebox[\textwidth]{%
        \includegraphics[width=0.85\textwidth]{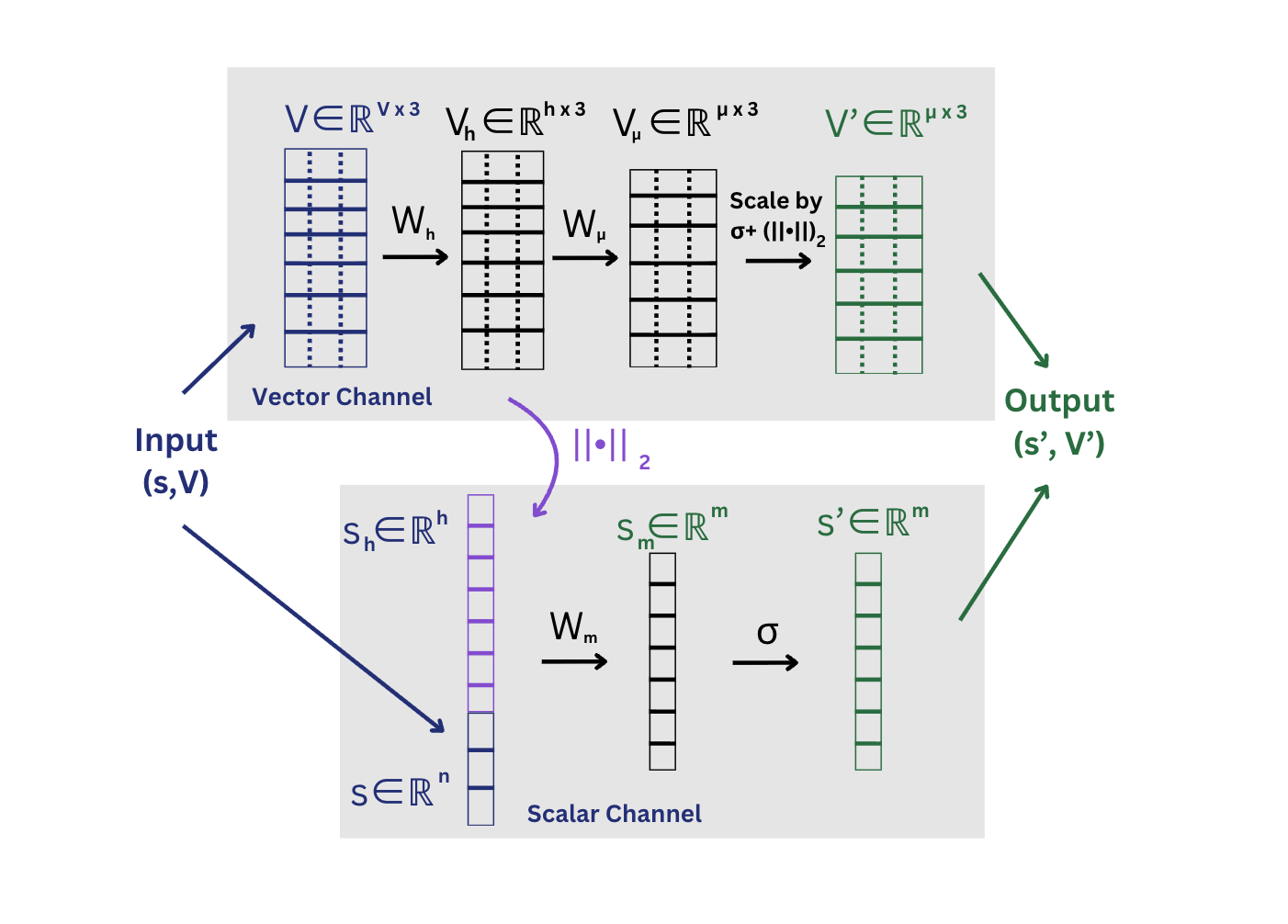}
    }
    \caption{GVP structure. (Graph recreated by the student researcher using Canva based on \cite{jing2020learning})}
    \label{fig:GVP-GCN}
\end{figure}

The GVP module first applies nonlinear transformations to the features while ensuring geometric consistency. Specifically, ReLU activations are applied to the scalar features, and a gating mechanism is applied to the vector features, ensuring rotational invariance. This design allows the model to refine nucleotide representations by incorporating local information from neighboring nucleotides. After the feature transformation, the GVP graph convolution (GVP) aggregates information from each nucleotide’s nearest neighbors, using the previously computed scalar and vector features $(\mathbf{s, V})$ to update the features $(\mathbf{s', V'})$ based on local geometric relationships.

To further enhance the model's ability to capture complex inter-nucleotide dependencies, random edge features are introduced at each convolution step. This helps the model learn additional nuances between the nucleotides. The resulting GVP layer performs the graph convolution as described in Algorithm 1, refining the features with each layer.

\begin{align}
    \textbf{GVP} (\mathbf{s, V}) = (\mathbf{s', V'})
\end{align}
\vspace{3pt} 
\hrule
\vspace{3pt} 
\noindent 
\textbf{Algorithm 1:} Geometric Vector Perceptron Algorithm from \cite{jing2020learning}
\vspace{5pt}
\hrule
\vspace{5pt} 

\noindent
\textbf{Input:} Scalar and vector features $(\mathbf{s, V}) \in \mathbb{R}^n \times \mathbb{R}^{\nu \times 3}$\\
\textbf{Output:} Scalar and vector features $(\mathbf{s', V'}) \in \mathbb{R}^m \times \mathbb{R}^{\mu \times 3}$.

\noindent
$h \gets \max(\nu, \mu)$

\noindent
\textbf{GVP:}
\begin{align}
    V_h &\gets W_h V &&\in \mathbb{R}^{h \times 3} \\
    V_\mu &\gets W_\mu V_h &&\in \mathbb{R}^{\mu \times 3} \\
    s_h &\gets \|V_h\|_2 \text{ (row-wise)} &&\in \mathbb{R}^{h} \\
    v_\mu &\gets \|V_\mu\|_2 \text{ (row-wise)} &&\in \mathbb{R}^{\mu} \\
    s_{h+n} &\gets \text{concat}(s_h, s) &&\in \mathbb{R}^{h+n} \\
    s_m &\gets W_m s_{h+n} + b &&\in \mathbb{R}^{m} \\
    s' &\gets \sigma(s_m) &&\in \mathbb{R}^{m} \\
    V' &\gets \sigma^+(v_\mu) \odot V_\mu \text{ (row-wise multiplication)} &&\in \mathbb{R}^{\mu \times 3}
\end{align}

\noindent
\textbf{return} $(\mathbf{s', V'})$
\vspace{5pt}
\hrule
\vspace{5pt} 
\vspace{10pt} 

To increase the model’s ability to capture hierarchical and local patterns, I use six GVP convolutional layers. These layers progressively expand the feature set with additional learned representations before condensing them into a lower-dimensional space while retaining the critical structural information necessary for RNA folding prediction. The output of the GVP are the post-processing high-level features from key atoms.

\subsection{Transformer}
Then, the new scalar and vector features are fed through a transformer encoder. I convert token indices from the RNA sequence input \(X\) to embeddings. The sequence length features go through a fully connected layer and are then added to each token embedding. Then, I add positional encoding to each embedding to ensure the transformer can capture local and global dependencies, making the final input embedding as below.
\begin{equation}
    H = \text{token\_embeddings} + \text{position\_embeddings}
\end{equation}

The inputs are then run through a transformer. I have a total of 6 layers, with 8 attention heads in each. A diagram of my transformer encoder-decoder is below in Figure~\ref{fig:transformer}.

\begin{figure}[!ht]

     \makebox[\textwidth]{%
        \includegraphics[width=1\textwidth]{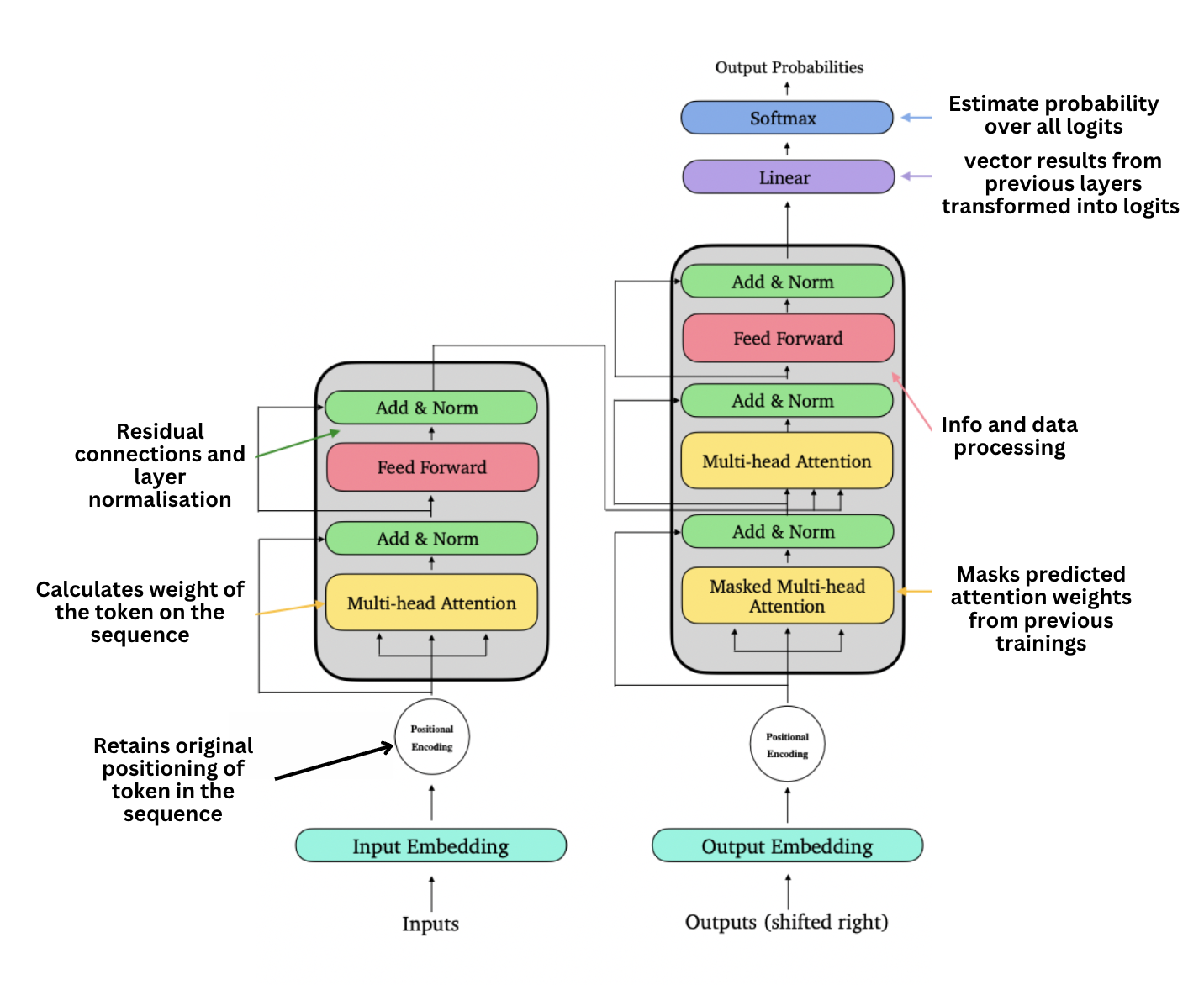}
    }
    \caption{Transformer structure. (Graph created by the student researcher using Canva based on prior research \cite{yao2024novel})}
    \label{fig:transformer}
\end{figure}

The transformer processes the embeddings, with the model taking the output of the current sequence as the input for the decoder, and then autoregressively outputs the predicted sequence. For every autoregressive generation of a fixed sequence length, a forward pass through the transformer decoder is conducted for the sequence length feature. The model will output probability distributions over the vocabulary for every position in the sequence, generating the sequence one token at a time by evaluating the previous token and selecting the most probable token as the next token in the sequence. The newly predicted token is then appended to the generated sequence. The decoder will continuously generate tokens for the sequence until an `end' token is generated or there are 50 tokens in the sequence. 

\subsection{Model Training and Evaluation}
I train my GVP-Transformer model end-to-end on a curated RNA dataset, splitting it into training, validation, and test sets. All RNA structures are preprocessed to form graphs suitable for the GVP encoder, and the corresponding sequences are tokenized to produce indices for the Transformer. Training is conducted on two NVIDIA GeForce RTX 4090 GPUs, allowing for efficient parallelization of both the GVP and Transformer components. I optimize the model parameters using the Adam optimizer with an initial learning rate of $1 \times 10^{-4}$, a batch size of 16, and a weight decay coefficient of $5 \times 10^{-5}$ to control overfitting. Dropout at a rate of 0.2 is applied to the Transformer layers, and gradient clipping is set to a global norm of 1.0 to stabilize updates. The model is trained for up to 80 epochs, with early stopping triggered if the validation loss fails to improve for 5 consecutive epochs. I use cross-entropy as the loss function.

I evaluate my RNA sequence design using two key performance metrics: recovery rate and structural recovery. These metrics were calculated based on the comparison between the designed RNA sequences and their natural counterparts, as well as the structural prediction of the designed sequences. For each metric, a higher score indicates a superior outcome, with 0.5 considered a high score. 

\begin{itemize}
    \item The recovery rate (\(R\)) is calculated by comparing the designed RNA sequence with the natural RNA sequence. The recovery rate is determined by the formula:

\begin{align}
R = \frac{\text{Number of correctly recovered nucleotides}}{\text{Total number of nucleotides}} \times 100
\end{align}

Where the "number of correctly recovered nucleotides" refers to the nucleotides in the designed RNA sequence that match exactly with the natural RNA sequence. The "total number of nucleotides" refers to the length of the RNA sequence.

\item \textbf{Structural recovery.}
I evaluated the structural similarity between the predicted structure ($S_{\text{pred}}$) and the experimental structure ($S_{\text{exp}}$) using the TM-score \cite{zhang2004scoring}. After optimal rigid-body superposition, the TM-score is

\begin{equation}
\mathrm{TM} = \max \left[ \frac{1}{L_{\text{ref}}} \sum_{i=1}^{L_{\text{ali}}} \frac{1}{1 + \left( \frac{d_i}{d_0(L_{\text{ref}})} \right)^2 } \right],
\end{equation}

where $d_i$ is the distance between the $i$-th pair of aligned residues in $S_{\text{pred}}$ and $S_{\text{exp}}$, $L_{\text{ref}}$ is the length of the reference structure, $L_{\text{ali}}$ is the number of aligned residue pairs, and $d_0(\cdot)$ is a length-dependent scale (e.g., $d_0(L)=1.24\sqrt[3]{L-15}-1.8$ in \AA). Higher TM-scores indicate greater structural similarity.

To obtain $S_{\text{pred}}$, I folded the designed sequences and then computed TM-scores to the corresponding experimentally solved structures. 
\end{itemize}

\section{Results}

\subsection{Model Performance and Benchmarking}
Table 1 shows the comparison of recovery rates and TM-scores for RNA sequence recovery and structural prediction. My model outperforms the other models in all areas consistently, achieving a higher recovery rate and TM-score for both standard benchmark and RNA-puzzles.
\begin{table}[ht]
    \centering
    \begin{tabular}{l|c|c|c|c}
        \hline
        Methods/Metrics & \multicolumn{2}{c|}{Recovery Rate} & \multicolumn{2}{c}{TM-score} \\
        \hline
        & Standard  & RNA-Puzzles & Standard  & RNA-Puzzles \\
        \hline
        \textbf{Ours}    & \textbf{0.413} & \textbf{0.481} & \textbf{0.281} & \textbf{0.332} \\
        RDesign & 0.328 & 0.403 & 0.252 & 0.272 \\
        Ribologic \cite{wu2019automated} & 0.142 & 0.242 & 0.190 & 0.250 \\
        Learna \cite{runge2024machine} & 0.213 & 0.293 & 0.224 & 0.194 \\
        \hline
    \end{tabular}
    \caption{Comparison of recovery rates and TM-scores for RNA sequence recovery and structural prediction. The first two columns correspond to results from the standard benchmark, while the last two columns show the results from the RNA-Puzzles dataset.}
    \label{tab:merged_benchmark_results}
\end{table}

To evaluate the performance of my RNA sequence design, I conducted a 10-fold cross-validation. This process involves dividing the dataset into 10 subsets, where 9 subsets are used for training and the remaining subset is used for testing. The procedure is repeated 10 times, ensuring that each subset serves as the test set once. The results are then averaged over all 10 folds to provide a robust measure of performance. In my comparison, I first downgraded my 3D RNA structure into 2D and then applied 2D sequence design methods. This approach allowed me to compare my method with other established 2D design methods. For structural comparison, I used AlphaFold3 to fold back the designed sequences. The recovery rate and TM-score were used as key performance metrics. The recovery rate measures how well the designed sequence matches the natural RNA sequence, while the TM-score evaluates the similarity between the predicted and experimental structures. In my experiments, based on 80\% sequence similarity cutoff for train-test split, the methods were compared. The results of my standard benchmark, as shown in Table \ref{tab:merged_benchmark_results}, demonstrate that my approach outperforms other methods such as Ribologic and Learna in both recovery rate and TM-score.

\begin{figure}[!ht]
    \centering

    \includegraphics[width=0.98\textwidth]{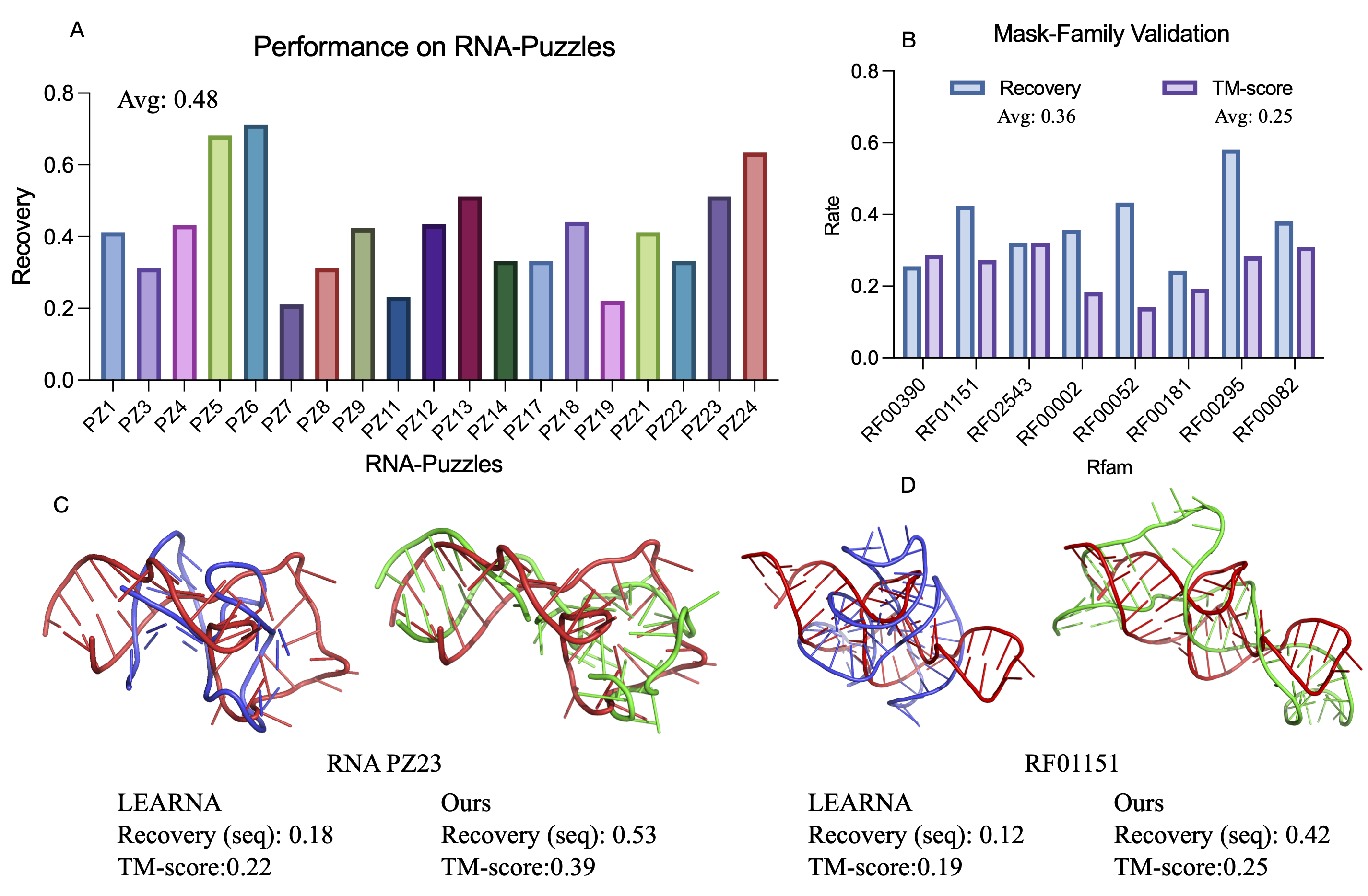}
    \caption{Results on RNA-Puzzles and Mask-Family. (A) Bar chart of the recovery rate for each RNA-Puzzle ID using my method. (B) Bar chart of the validation results for the Mask-Family task, reporting both recovery rate and fold-back TM-score for each family. (C) Comparison of RNA design outcomes for PZ23, contrasting LEARNA's predictions (blue structure) with ours (green), both folded using AlphaFold3 and overlaid with the ground-truth structure (red). (D) The same comparison for the RF01151 family. (Figures A and B are created by the student researcher using GraphPad Prism and Figures C and D are created by the student researcher using PyMOL)}
    \label{fig:IHAD}
\end{figure}

\subsection{Mask-Family Validation}

To further assess the performance of my RNA sequence design, I conducted a cross-family validation using Rfam annotations. In this validation, I split my dataset into training and test sets using a leave-one-out strategy. Specifically, for each family in the dataset, I trained the model on all families except for one, which was held out as the test set. This process was repeated for each family, ensuring that every family was tested independently. This approach provides a more robust evaluation of how well the model generalizes across different RNA families.

I used the curated PDB dataset from the previous experiment, along with family annotations provided by Rfam. The Rfam annotations categorize RNA families, allowing me to validate the model's ability to predict sequences and structures for RNA families it has not seen before. My method received an average Recovery Rate of 0.36 and average TM-score of 0.254 for cross-family results, showing robust results across diverse families (Figure \ref{fig:IHAD}B). I compared this to Learna's sequence generalizability of 0.155 and my model achieved higher results. 

\subsection{RNA-puzzle Evaluation}

In addition to cross-family validation, I also tested my model on open-wide challenges such as RNA-Puzzles, a renowned benchmark for RNA sequence design and structure prediction. For this evaluation, I filtered out sequences with a sequence similarity greater than 80\% from my training set to ensure that the test sequences were sufficiently distinct from those used during training. This approach is designed to test the model’s generalization ability on new and diverse RNA structures. I trained and tested my model on the RNA-Puzzles dataset and evaluated its performance using the recovery rate and TM-score metrics. The recovery rate measures the accuracy of the designed RNA sequences when compared to the natural sequences, while the TM-score evaluates the similarity between the predicted and experimental RNA structures. 

The results of this evaluation (Table \ref{tab:merged_benchmark_results}, Figure \ref{fig:IHAD}A) demonstrate that my method outperforms other approaches, such as RDesign, Ribologic, and Learna, in both recovery rate and TM-score. Specifically, my method achieved a recovery rate of \textbf{0.481} and a TM-score of \textbf{0.332}, indicating superior performance in both sequence recovery and structural prediction accuracy.

\section{Discussion}

Compared to conventional RNA inverse folding methods, my model demonstrates substantial improvements across multiple evaluation benchmarks. On both the standard benchmark set and the RNA-Puzzles challenge, it consistently outperforms existing approaches in terms of sequence recovery and structural fidelity. For RNA-Puzzles, my model achieves a recovery rate of 0.481 and a TM-score of 0.332. Given that TM-scores are calculated by refolding the generated sequences using AlphaFold3 and comparing the resulting structures with experimentally determined ones, it is important to note that AlphaFold3 itself typically yields an average TM-score of approx. 0.5. Thus, a TM-score of 0.332, starting from de novo generated sequences, is high and demonstrates the structural validity of my designs. Additionally, the recovery rate, which evaluates exact sequence-level matches, surpasses the TM-score, further underscoring the precision of my model in generating structurally informed sequences. My method also excels in Mask-Family validation, achieving a sequence similarity score of 0.42 for RF01151. In this setting, scores above 0.4 are considered statistically significant, highlighting the model’s strong generalization capacity across different RNA families.

When benchmarked against existing models, my approach demonstrates clear superiority. For RNA-Puzzles, Ribologic reports a recovery rate and TM-score of 0.242 and 0.250, respectively, while Learna performs even lower at 0.293 and 0.194. RDesign, despite achieving a relatively high recovery rate of 0.403, lags in structural accuracy with a TM-score of only 0.272. In contrast, my model achieves high performance across both metrics, reflecting its balanced capability in accurate sequence generation and structural realization. Similarly, in the standard benchmark, my model attains a recovery rate of 0.413 and a TM-score of 0.281. These results outperform RDesign (recovery rate 0.328; TM score 0.252), Ribologic (recovery rate 0.142; TM-score 0.190), and Learna (recovery rate 0.213; TM-score 0.224). Notably, my model improves upon Ribologic’s recovery rate by nearly fourfold and more than doubles that of Learna. These comprehensive gains across both sequence- and structure-level metrics firmly establish the superiority of my framework and mark an advancement toward reliable, high-accuracy RNA design. 

Despite its strong performance, my model has several limitations. It struggles with highly complex RNA structures, such as those containing long-range interactions, pseudoknots, and multi-loop junctions, although geometric features help mitigate this challenge. The model is limited to single RNA design and does not account for co-folding or binding constraints in RNA complexes. Additionally, it predicts only the most stable conformation, ignoring alternative functional states common in riboswitches and ribozymes. Finally, the scarcity of high-quality RNA structural data hampers training, underscoring the need for expanded datasets to further enhance model performance.

My new method of computational RNA inverse-folding has a wide range of applications. Most notably, its ability to encode precise protein-binding RNA architectures will allow me to engineer RNAs that recruit specific RNA-binding proteins to targeted transcripts, enabling post-transcriptional regulation with high molecular specificity. This same framework can be used to create therapeutic aptamers—structured RNAs that tightly bind and neutralize disease-associated proteins—through controlled presentation of binding loops and tertiary contacts. By starting from the desired structure and going backward to the nucleotide sequence, I can efficiently explore a much broader region of the design space, identifying sequence variants that preserve the functional fold, thus designing its interaction capabilities. This work highlights the transformative potential of machine learning in RNA research and positions AI-driven methods as foundational tools for the next generation of molecular design.

\section{Conclusion}

In this research, I identified a significant problem in the field of RNA sequence discovery and created and applied a new framework to provide a solution for the problem. My created framework generates results that surpass the capabilities of all compared RNA prediction models with higher recovery rates and TM-scores, demonstrating superior sequence reconstruction capabilities and structural prediction accuracy. I also made my code open-source so that the public has free access to my created model.\footnote{\url{https://github.com/Annabelleyao/GVP-Transformer-for-RNA-Design}} Prior to the rise of machine learning, RNA design relied heavily on dynamic programming tools such as ViennaRNA and RNAfold, which struggle with scalability due to the exponential growth of possible structures with RNA length. For each RNA of length \( n \), there are  \(n^{-3/2} \times 1.8^n\) possible structures, each taking a large amount of time to design. Experimental design was often a slow, trial-and-error process, taking weeks to months. In contrast, my model can generate structurally valid sequence candidates within minutes, dramatically accelerating RNA design workflows. By reducing both design time and computational burden, my framework paves the way for rapid innovation in RNA-based therapeutics, vaccine development, gene regulation, and synthetic biology. 

\newpage

\bibliography{mybib}
\addcontentsline{toc}{section}{References}

\end{document}